\documentclass[11pt]{article}
\usepackage[]{acl2023}%
\usepackage[most]{tcolorbox}
\usepackage{tabularx}
\usepackage{newtxtext}
\usepackage{latexsym}
\usepackage{graphicx}
\usepackage{array}
\usepackage{amsmath}
\usepackage{amsfonts}
\usepackage{booktabs}
\usepackage{nameref}
\usepackage{comment}
\usepackage{cleveref}
\usepackage{multirow}
\usepackage{caption}
\usepackage{placeins}
\usepackage{xspace}
\usepackage{wrapfig}
\usepackage{subcaption}
\usepackage{enumitem}
\usepackage[dvipsnames]{xcolor}
\usepackage{todonotes}
\usepackage{ulem}

\newcommand{\ARR}[1]{{ \textcolor{red}{#1}}}

\usepackage{xcolor} 

\usepackage{microtype}

\title{Evaluating Context Sensitivity of Collaboration Behavior \\in Multi Party Simulation Frameworks}

\author{
    \textbf{Anvesh Rao Vijjini}\textsuperscript{1,$\dagger$} \quad
    \textbf{Sheshera Mysore}\textsuperscript{2} \quad
    \textbf{Pegah Jandaghi}\textsuperscript{3,$\dagger$} \\
    \textbf{Ashish Sharma}\textsuperscript{2} \quad
    \textbf{Snigdha Chaturvedi}\textsuperscript{1} \quad
    \textbf{Bahareh Sarrafzadeh}\textsuperscript{2} \\[2pt]
    \textsuperscript{1} University of North Carolina at Chapel Hill \qquad
    \textsuperscript{2}Microsoft \\
    \textsuperscript{3}University of Southern California \\
    \texttt{\{anvesh,snigdha\}@cs.unc.edu jandaghi@usc.edu} \\
    \texttt{\{sharma.ashish,basarraf\}@microsoft.com} \quad \texttt{smysore@iesl.cs.umass.edu}
}

\date{}

\begin{document}

\maketitle

\begingroup
\renewcommand{\thefootnote}{\fnsymbol{footnote}}
\footnotetext[2]{Work conducted during internships at Microsoft.}
\endgroup
\begin{abstract}

Group Social Behavior is increasingly used as a lens to study Large language model (LLM) simulations to aid model training and evaluation. However, little prior work has examined how contextual factors such as team size and goal type manifest in long running simulations. To overcome this gap, we conduct a controlled, full-factorial study examining how contextual factors affect collaboration behavior across multi-party simulation frameworks. Using well established social science taxonomies \cite{bales1950interaction,meier2007rating}, we quantify context sensitivity of collaboration behavior in simulated teams. Our analyses reveal significant effects of contextual factors such as simulators consistently favor smooth collaboration over disagreement, negotiation pressure, and coordination breakdown. These findings provide the first systematic evidence of how contextual factors influence emergent collaboration in LLM based multi-party simulation frameworks, offering both guidance and empirical results for the use of such simulations.
\end{abstract}

\section{Introduction}
\label{sec:intro}

%

Multi-party simulations have long been used to study the emergence of complex social and behavioral phenomena in artificial societies \cite{epstein1996growing,gilbert2018simulating}. Recent advances in large language models (LLMs) have renewed interest in this paradigm by enabling LLM-based simulators that communicate in natural language, adopt roles, and participate in open ended interactions \cite{park2023generative, li2023camel, zhou2024sotopia}. Such simulations provide scalable environments for studying behavior without direct human experimentation \cite{argyle2023outofone}, estimating the effects of environmental or policy changes before deployment \cite{park2024sim1k, sreedhar2025prosocial}, generating synthetic interaction data for downstream training \cite{abdullin2023synthetic, tang2025matrix}, or exploring emergent phenomena such as coordination, cooperation, and conflict \cite{piatti2024govsim, akata2025games}.






However, the usefulness of an LLM-based simulator depends on whether its behavior changes meaningfully when \textit{contextual factors} such as team composition and size are changed. Concurrent works on user simulation investigate this question. \citet{mehri2026measuring} argue that while simulators can generate human-like responses, it remains unclear whether they capture the broad distribution of user behaviors. Similarly, \citet{suh2026quantifying} argue that simulator quality should be measured by downstream utility, such as whether assistants trained with a simulator perform better. Other works on LLM based social simulation have explored role playing, emergent communication, negotiation, and coordination among agents \cite{park2023generative, li2023camel, zhou2024sotopia, abdelnabi2024llmnegotiation}. 

 Furthermore, LLM-based simulations are going beyond dyadic user assistant settings to multi-party settings in which AI systems participate in, mediate, or simulate group processes. \citet{liu2025innerthoughts} study proactive conversational agents that decide when to contribute in multi-party simulation. \citet{lee2025amplifying} propose AI devil's advocate agents that surface minority perspectives in group decision making. \citet{wang2026seesawbot} propose Slack-based mediators that move between private and shared team channels to support team dynamics. \citet{liu2025promediate} study proactive mediator agents for multi topic, multi party negotiations . 
Increasingly, AI products in industry are also focusing on assistance in multi-party settings, such as shared ChatGPT group chats, Slack-integrated agents~\citep{openai2025groupchats, manus2025slack, granola2025teams, dust2026multiplayer}. For all these applications, it becomes important to evaluate simulators for multi-party simulations.


These developments leave two gaps in how LLM based multi party simulators are evaluated. First, existing evaluations often focus on surface measures such as task success, coherence, or generation quality. Second, these works typically study a specific agent design, interaction technique, or task setting, without jointly examining how multiple aspects of the group context and simulator design shape behavior. We address these gaps by evaluating simulator sensitivity: \textit{whether controlled changes in contextual factors provided to the simulation such as team composition, team size, goal type, and simulation framework; lead to systematic and measurable changes in how simulated groups communicate, coordinate, build consensus, and resolve disagreement} \cite{hinde1987individuals, mcgrath1984groups}.

 We focus on three key contextual factors that are central to group interaction grounded in cognitive science and group interaction literature: \textit{(i)} team composition, comparing uniform and diverse expertise \cite{stasser1985pooling,jehn1995multidimensional}, \textit{(ii)} team size which can be 3, 5, 7 or 9 members \cite{mcgrath1991time,latane1979social}, and \textit{(iii)} goal type, using four goal categories such as generating ideas, choosing among options, negotiating disagreement, and executing a plan \cite{mcgrath1984groups,amabile1996creativity}. We evaluate these factors across three LLM based multi party simulation frameworks: Sotopia \cite{zhou2024sotopia}, TinyTroupe \cite{salem2025tinytroupe}, and ``Scripted''. In total, our evaluation spans 96 controlled settings, covering 2 team compositions, 4 team sizes, 4 goal types, and 3 frameworks, with up to 10 scenarios generated for each setting. To measure collaboration behavior, we adapt two established coding frameworks from collaboration research. Bales' Interaction Process Analysis provides an utterance-level taxonomy for measuring local interaction patterns, such as information sharing, suggestions, agreement, and disagreement (\textit{Bales IPA}) \cite{bales1950interaction}. Meier's Quality of Collaboration framework complements this by capturing higher-level properties of the full interaction, such as information pooling, task coordination, consensus building, and mutual understanding (\textit{Meier's collaboration quality dimensions}) \cite{meier2007rating}. Together, these frameworks allow us to quantify both local conversational behavior and broader collaborative quality. We use an LLM judge to annotate simulated interactions along these dimensions, and use statistical testing to examine whether each contextual factor produces significant differences in collaboration behavior while averaging over the other factors.




Our results show that LLM based multi party simulation frameworks are sensitive to contextual factors, but in an uneven manner. Goal type produces the clearest behavioral differences, while team size effects vary across frameworks. Across settings, simulators reliably produce task-oriented collaboration, such as information sharing, coordination, and agreement, but underrepresent disagreement, negotiation pressure, and coordination breakdown. In Appendix \ref{sec:related} we discuss more related works. 

\begin{figure*}
    \centering
    \includegraphics[width=\linewidth]{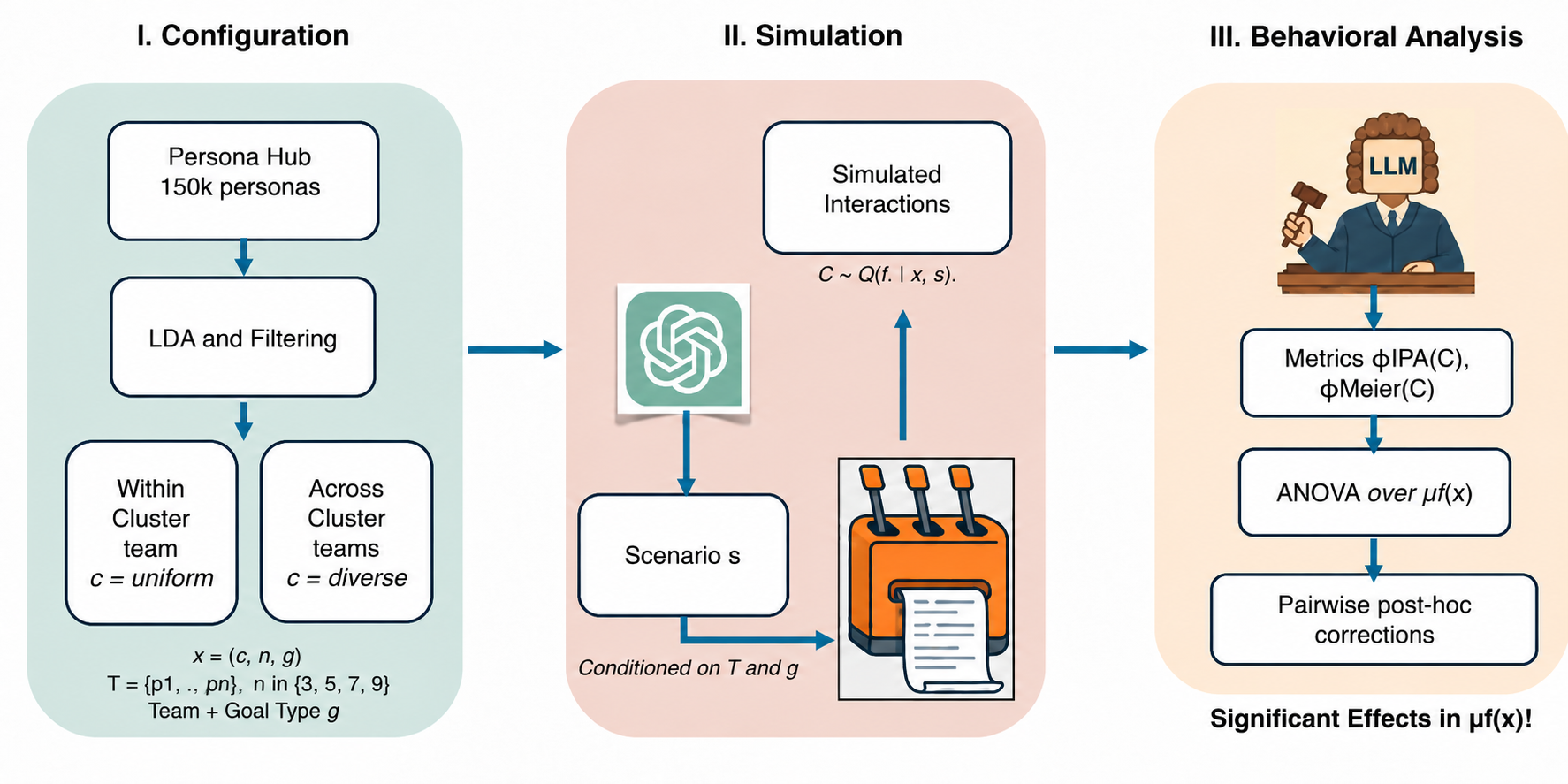}
    \caption{Pipeline for evaluating context sensitivity in LLM based multi party simulators. Configurations vary team composition, team size, and goal type, simulations generate conversations. Finally, behavioral analysis maps them to Bales IPA and Meier metrics for statistical testing.}
    \label{fig:pipeline}
\end{figure*}

\section{Problem Statement}
\label{sec:problem}
We study the problem of evaluating whether LLM based multi party simulation frameworks produce collaboration behavior that is sensitive to controlled changes in the simulation. In other words, we ask whether changing team size, team composition, or goal type changes the measured collaboration behavior of the simulated group. A useful simulator should not only generate coherent conversations, but should also produce measurably different collaboration behaviors when relevant contextual factors are changed.
Let $f \in \mathcal{F}$ denote a simulation framework, where $\mathcal{F}$ includes Sotopia, TinyTroupe, and  ``Scripted''. Each simulation is conditioned on a configuration
\[
x = (c, n, g),
\]
where $c$ denotes team composition, $n$ denotes team size, and $g$ denotes goal type. In our study, team composition is either uniform or diverse, team size ranges from three to nine agents, and goal type follows four categories of group activity: generating, choosing, negotiating, and executing.

A configuration $x$ specifies a team of agents
$T = \{p_1, p_2, \ldots, p_n\}$,
where each $p_i$ is a persona sampled from a
persona pool according to composition $c$ and
team size $n$. Given a sampled team $T$ and
goal type $g$, we generate a concrete collaborative
\textit{scenario} $s$ conditioned on $(T,g)$. For example,
a team comprising an engineer, lawyer, and artist
assigned a negotiation goal might be tasked with
reconciling technical, legal, and aesthetic
requirements for a public building. Thus, $s$
represents a specific task description rather than
a set of agents or goal types. The simulator
framework $f$ then generates a multi-turn group
interaction

\[
C = ((r_1, y_1), (r_2, y_2), \ldots, (r_m, y_m)),
\]
where $r_t$ is the speaking agent at turn $t$ and $y_t$ is the corresponding utterance.

Each framework induces a conditional distribution over conversations:
\[
Q_f(C \mid x, s).
\]
Our goal is to characterize how this induced distribution changes as the configuration $x$ changes. Since the full conversation distribution is not directly interpretable, we map each conversation to a set of collaboration behavior measurements:
\[
\phi(C) = [\phi_{\text{IPA}}(C), \phi_{\text{Meier}}(C)].
\]
Here, $\phi_{\text{IPA}}(C)$ denotes the distribution of utterances across Bales' Interaction Process Analysis categories, and $\phi_{\text{Meier}}(C)$ denotes the presence of collaboration quality dimensions such as information pooling, communication quality, consensus building, task management, mutual understanding, and argumentation.

For each framework $f$ and configuration $x$, we estimate the expected collaboration behaviours:
\[
\mu_f(x) = \mathbb{E}_{C \sim Q_f(\cdot \mid x, s)}[\phi(C)].
\]
A simulator is context sensitive if changes in $x$ lead to statistically significant changes in $\mu_f(x)$.

This formulation leads to three evaluation questions. First, what baseline collaboration behaviors do different simulation frameworks produce? Second, which contextual factors significantly shape collaboration behavior? Third, do different frameworks exhibit different forms of sensitivity to the same contextual interventions? These questions allow us to evaluate multi agent simulators not only by whether their conversations appear plausible, but by whether their behavior changes in systematic and measurable ways under controlled experimental variation.


\section{Experiments}
We operationalize the formulation in Section~\ref{sec:problem} through a full-factorial experiment over team composition, team size, goal type, and simulation framework. This section describes the concrete implementation choices: how personas are selected and grouped, how scenarios are generated, how simulations are run, and how collaboration behavior is annotated and statistically tested.

\begin{table*}[t]
\centering
\small
\setlength{\tabcolsep}{5pt}
\renewcommand{\arraystretch}{1.15}
\begin{tabularx}{\linewidth}{
    @{}l
    >{\raggedright\arraybackslash}X
    >{\raggedright\arraybackslash}X@{}
}
\toprule
\textbf{Goal Type ($g$)}  &
\textbf{Example Scenario ($s$)} &
\textbf{Example Team ($T$)} \\
\midrule

Executing &
Organize a community event on property rights,
groundwater management, and local cultural
significance. &
\textit{3 agents:} Estate agent, hydrogeologist,
literature student. \\

\addlinespace

Choosing &
Analyze vaccine hesitancy in a multicultural
neighborhood and choose an effective
communication strategy. &
\textit{9 agents:} Medical social worker,
journalism/media voices, counseling psychologist,
surgeon, small-business owner, and 4 others. \\

\addlinespace

Executing &
Organize a live community fair combining
sports, music, and fashion while coordinating
roles and event flow. &
\textit{7 agents:} Libertarian activist,
violinist/UFC fan, prison officer,
volleyball coach, streetwear enthusiast,
and 2 others. \\

\addlinespace

Generating &
Brainstorm an AI-driven language-learning
intervention for impulse control and
cognitive function in mental health. &
\textit{7 agents:} Psychiatrists, AI app developer,
environmental scientist, retailer, patient,
and 2 others. \\

\addlinespace

Negotiating &
Negotiate a fair allocation of limited funding
across health, business, culture, music,
literature, sports, and recovery projects. &
\textit{7 agents:} Occupational health nurse,
business partner, doctor, musician,
industry veteran, and 2 others. \\

\bottomrule
\end{tabularx}
\caption{Examples of collaborative scenarios and
teams generated under different goal types.
Each scenario $s$ specifies a task conditioned
on the sampled team $T$ and goal type $g$.
For larger teams, only a subset of personas
is shown.}
\label{tab:scenario_examples}
\end{table*}
\subsection{Configuration and Simulation Frameworks}

We evaluate collaboration behavior across three simulation frameworks: \textbf{Sotopia} \cite{zhou2024sotopia}, \textbf{TinyTroupe} \cite{salem2025tinytroupe}, and a \textbf{``Scripted''} baseline.  
Sotopia provides structured, goal-oriented social interactions between LLM-based simulators, allowing for rich negotiation and multi-turn collaboration. TinyTroupe, in contrast, is a lightweight framework that supports flexible agent configurations and conversational dynamics with minimal role constraints.  
The ``Scripted'' framework consists of template-driven interactions in which dialogue roles and turn-taking are partially pre-defined, producing more deterministic but less adaptive exchanges.  

Overall, there is a need for cross-framework evaluation for realism and evaluating sensitivity of simulation frameworks to contextual factors. This is further backed by prior work \cite{zhou2024real} has shown that simulation design can significantly influence observed social-interaction outcomes, so there is a need for cross-framework evaluation for realism and sensitivity testing. 

 All simulations in our study use \textbf{GPT-4.1} for agent generation and \textbf{GPT-5-chat} for evaluation. We fix the models across all conditions to isolate contextual and framework effects, and to keep the experiment computationally tractable.

\subsection{Configuration Design and Simulation}

We adopt a controlled design over three contextual factors: team composition, team size, and goal type. These factors define the \textit{configuration} stage of the pipeline (Figure~\ref{fig:pipeline}). Each configuration $x = {c,n,g}$ specifies a team and a goal type, which are then instantiated into concrete scenarios and simulated across frameworks.

\paragraph{Team composition.}
We obtain a pool of 150K personas from PersonaHub
\citep{ge2024scaling} and apply Latent Dirichlet
Allocation (LDA) \cite{blei2003latent} to identify topic-based clusters based
on persona descriptions. Each persona is assigned
to the cluster corresponding to its dominant topic.
We then filter out non-collaborative clusters, such
as family relationships or purely
personal roles (``a young couple who are
adventurous eaters'' or ``a single mother looking
for kid-friendly party options.'')

Using the remaining clusters, we construct two types of teams:
\begin{itemize}[topsep=1pt,noitemsep,leftmargin=*]
    \item \textbf{Uniform:} all members come from the same cluster, such as teacher, principal, and counselor.
    \item \textbf{Diverse:} members span distinct clusters, such as lawyer, engineer, and teacher.
\end{itemize}

\paragraph{Team size.}
We vary team size $n \in \{3, 5, 7, 9\}$ to capture coordination effects across small to moderately large groups.

\paragraph{Goal type.}
We adopt the four goal categories proposed by \cite{mcgrath1984groups}, which capture major functional modes of group interaction:
\begin{enumerate}[topsep=1pt,noitemsep,leftmargin=*]
    \item \textbf{Generating:} idea creation and brainstorming.
    \item \textbf{Choosing:} evaluation and decision-making.
    \item \textbf{Negotiating:} conflict resolution and alignment.
    \item \textbf{Executing:} performance-oriented coordination.
\end{enumerate}

For each configuration $x=(c,n,g)$, we sample a team
$T$ according to composition $c$ and size $n$,
and generate a concrete collaborative scenario $s$
conditioned on $(T,g)$. Here, $s$ specifies a
realistic task that the team would naturally perform,
such as teachers brainstorming a new curriculum
for a generating goal. To improve scenario validity,
the model may skip generation if no plausible scenario
exists, such as if the team combination is just too implausible to create a scenario. We repeat this process to obtain up to 10 distinct scenarios per configuration $x$, accounting for scenario-level
variability. 

For each framework $f$ (Sotopia, TinyTroupe, and ``Scripted'') and configuration $x$, we
simulate conversations $C$ using instantiated teams
$T$ and collaborative scenarios $s$.

Each conversation is then passed to the behavioral
analysis stage of the pipeline. Overall, our full-factorial design spans
$2 \times 4 \times 4 = 32$ configurations per framework, with multiple scenarios generated for each configuration. Further details on counts are provided in Appendix \ref{app:data_stats}. Table \ref{tab:scenario_examples} provides examples of $T$, $g$ and $s$.

\begin{table*}[t]
\centering
\small
\begin{tabular}{p{0.31\linewidth}p{0.61\linewidth}}
\toprule
\textbf{Bales IPA Category} & \textbf{Description} \\
\midrule
1. Shows solidarity & Raises others' status, offers help, or shows appreciation. \\
2. Shows tension release & Jokes, laughs, or expresses pleasure. \\
3. Agrees & Shows understanding, acceptance, or affirmation. \\
4. Gives suggestion & Proposes ideas, directions, or actions. \\
5. Gives opinion & Evaluates, analyzes, interprets, or expresses feelings. \\
6. Gives information & Shares facts, data, or personal experience. \\
7. Asks for information & Requests factual details or clarification. \\
8. Asks for opinion & Seeks evaluation or perspective. \\
9. Asks for suggestion & Seeks proposals for action or decision. \\
10. Disagrees & Expresses dissent, doubt, or passive rejection. \\
11. Shows tension & Displays frustration, anxiety, or concern. \\
12. Shows antagonism & Criticizes, asserts dominance, or deflates others' status. \\
\bottomrule
\end{tabular}
\caption{Bales' Interaction Process Analysis categories used for utterance level annotation. The judge assigns one primary category to each utterance. We additionally allow a fallback ``13. None'' label when no IPA category applies, but exclude it from the core IPA schema.}
\label{tab:bales_ipa}
\end{table*}

\subsection{Collaboration Behavior Metrics}

After simulations, we quantify collaboration quality using two behavioural frameworks adapted from collaboration research. 

\paragraph{Bales’ Interaction Process Analysis (IPA).}

Each utterance is evaluated using \textbf{GPT-5-chat} acting as an \textit{LLM-as-a-Judge}, assigning exactly one of Bales’ 12 IPA categories. We normalize and aggregate the predictions to get  distribution of interaction labels for each conversation.

\paragraph{Meier’s Quality of Collaboration.} 

Conversations are segmented into snippets (5–10 turns), and each snippet is annotated for presence of defined dimensions: \textit{sustaining mutual understanding, dialogue management, information pooling, reaching consensus, task division, time management, technical coordination, reciprocal interaction, and individual task orientation}, (Table \ref{tab:meier_dimensions} for more details behind the dimensions). Since multiple dimensions can co-occur in a snippet, each snippet may receive multiple labels. We then compute the overall presence-percentage (0-100\%) of each dimension across the full conversation. The percentage indicates the proportion of conversation snippets that show the specific dimension.

We also do manual precision evaluation to find that 84.1\% of Bales predictions and 100\% of Meier predictions by the judge to be accurate on 50 utterances and 50 conversation snippets each (sampled uniformly across all labels/dimensions). This experiment is further discussed in Appendix \ref{app:manual}. Appendix \ref{app:full_prmopts} discusses the full prompts in implementing the judge. The details regarding Bales and Meier metrics are adapted from the original literature \cite{bales1950interaction,meier2007rating}.

\begin{table*}[t]
\centering
\small
\begin{tabular}{p{0.25\linewidth}p{0.65\linewidth}}
\toprule
\textbf{Meier Dimension} & \textbf{Description} \\
\midrule
1. Sustaining mutual understanding & Speakers make contributions understandable, explain terms, paraphrase, elicit feedback, or demonstrate understanding. \\
2. Dialogue management & Speakers maintain smooth communication flow, manage turn taking, signal handovers, or repair misunderstandings. \\
3. Information pooling & Participants collect solution relevant information, relate new information to existing facts, and clarify information needs. \\
4. Reaching consensus & Participants discuss options, weigh arguments, challenge ideas, and finalize decisions based on evidence. \\
5. Task division & Participants split work into subtasks, plan steps, match subtasks to expertise, and divide work fairly. \\
6. Time management & Participants monitor time, schedule phases, prioritize discussion, and finish without rushing. \\
7. Technical coordination & Participants coordinate tools, shared editors, document handling, app switching, or parallel writing tools. \\
8. Reciprocal interaction & Participants respect each other, encourage contributions, offer constructive critiques, and share decision making. \\
9. Individual task orientation & Participants actively pursue a good solution, stay task focused, and show interest in high quality work. \\
\bottomrule
\end{tabular}
\caption{Meier collaboration quality dimensions used for snippet level annotation. Multiple dimensions can apply to the same conversation snippet.}
\label{tab:meier_dimensions}
\end{table*}
\begin{figure*}
    \centering
    \includegraphics[width=\linewidth]{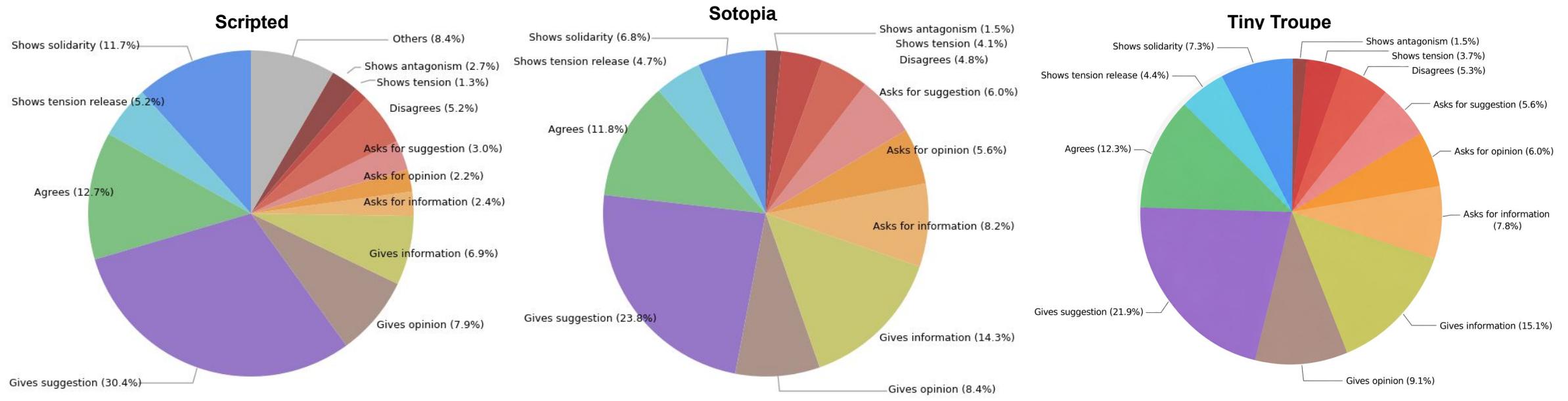}
    \caption{Overall Bales IPA distributions across Scripted, Sotopia, and TinyTroupe simulations. Simulated conversations are dominated by cooperative behaviors, while disagreement and tension remain rare.}
    \label{fig:overall_ipa_scripted}
\end{figure*}


These tables provide the coding schema for our behavioral measures. These serve as dependent variables in subsequent statistical testing of contextual factor effects.


\subsection{Significance Testing}
\label{sec:tukey}
To determine whether contextual factors significantly affect collaboration behavior, we perform factorial \textbf{ANOVA} \cite{st1989analysis} across all metrics and simulation frameworks. This allows us to identify both main effects and interaction effects of team size, team composition, and goal type.

Because multiple comparisons can introduce false positives \cite{tukey1949}, we apply post-hoc correction to adjust $p$-values. Because we test multiple metrics across several factors and frameworks, we use a conservative Bonferroni-style threshold of $p \leq 0.0005$. We report only effects that pass this threshold and treat all other effects as non-significant. For significant factors with more than two levels, we conduct post hoc pairwise comparisons to identify which levels differ.

Only statistically significant differences that survive correction are discussed across frameworks. All other effects are treated as non-significant.

\section{Results}

We organize our results around \textit{simulator sensitivity}: whether LLM based multi party simulation frameworks produce significant changes in collaboration behavior when we vary framework design, team size, and goal type. We ask which factors of the simulation steer behavior, and where the resulting behavior remains limited.

\subsection{Framework Design Shapes Baseline Collaboration Profiles}

We first examine the overall distribution of interaction behaviors across simulation frameworks using Bales' IPA categories (Figure~\ref{fig:overall_ipa_scripted}). This figure provides a baseline view of what simulated conversations are made of before conditioning on team size, composition, or goal type. Across all three frameworks, conversations are dominated by cooperative, task oriented behaviors, especially giving suggestions, giving information, giving opinions, and showing agreement. In contrast, behaviors associated with social friction, such as showing tension, antagonism, and disagreement, remain rare.

This result shows that LLM based multi party simulation frameworks reliably produce recognizable collaboration behavior. Agents exchange information, propose ideas, and move conversations toward agreement. However, the distribution also reveals an important limitation: the simulators produce cooperative structure more strongly than conflict, negotiation pressure, or disagreement. This matters because a simulator that produces fluent collaboration but rarely produces friction may be less useful for studying difficult group settings.

The differences across frameworks further show that simulator design is itself a source of behavioral variation. The ``Scripted'' framework produces the most concentrated pattern of suggestions and agreement. This is expected because a single centralized generation process can maintain global coherence across the full group interaction. Such coherence can make the conversation appear smooth, but it may also reduce agent independence and suppress emergent disagreement.

Sotopia and TinyTroupe distribute more behavior across information exchange, questions, suggestions, and opinions. Their more explicit agent abstractions make the conversations less centralized and more variable. At the same time, both still show limited disagreement and tension. Thus, framework design changes the baseline collaboration profile, but all three frameworks share a tendency toward polite and cooperative convergence.

Overall, Figure~\ref{fig:overall_ipa_scripted} shows that LLM based multi party simulation frameworks are not behaviorally neutral environments. Their design choices shape what kinds of collaboration appear, even before specific contextual factors are varied.

\subsection{Team Size Sensitivity Depends on Simulation Framework}

\begin{figure*}
    \centering
    \includegraphics[width=0.7\linewidth]{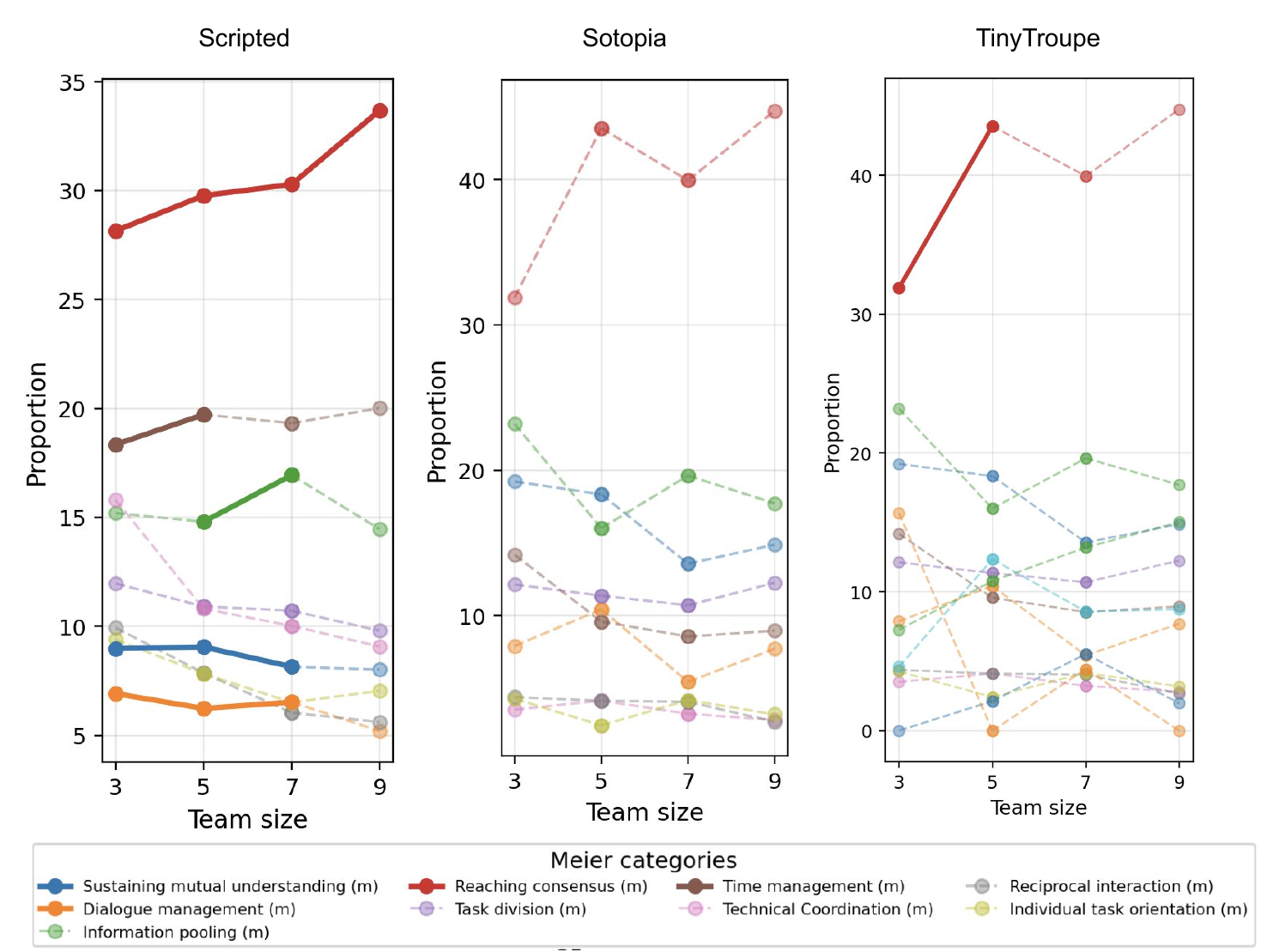}
    \caption{Effect of team size on Meier collaboration dimensions across frameworks. Bold lines indicate metrics with a statistically significant team size effect after correction.}
    \label{fig:meier_teamsize}
\end{figure*}

Figure~\ref{fig:meier_teamsize} shows how Meier collaboration metrics vary with team size across ``Scripted'', Sotopia, and TinyTroupe simulations. Bold lines indicate metrics for which team size has a statistically significant effect after correction, while non-bold lines are treated as descriptive trends.

The key finding is that team size does affect collaboration behavior, but the direction and strength of this effect depend on the simulation framework. In the ``Scripted'' framework, reaching consensus increases steadily as team size grows. This suggests that centralized generation can scale up apparent agreement as more participants are added. Rather than exposing coordination difficulty, the framework may produce consensus as a narrative pattern of group interaction.

Sotopia shows a less monotonic relationship between team size and collaboration metrics. Several metrics fluctuate across team sizes, suggesting that its turn based and environment mediated design introduces more interactional variability. This makes Sotopia less likely to produce smooth consensus by construction, although team size does not always lead to a clear degradation in coordination.

TinyTroupe shows stronger variation for some metrics, especially reaching consensus, but the trends are less stable across team sizes. Its persistent agent abstractions and memory may encourage explicit alignment among agents, while still allowing more variation than the ``Scripted'' setting.

Taken together, the team size results show that contextual sensitivity is framework dependent. The same intervention, increasing the number of participants, does not produce the same behavioral effect across simulators. This supports our broader framing: multi party simulators should be evaluated not only by the quality of their generated conversations, but by how their behavior changes under controlled contextual variation.

\subsection{Goal Type Produces the Clearest Behavioral Difference}

\begin{figure*}
    \centering
\includegraphics[width=0.9\linewidth]{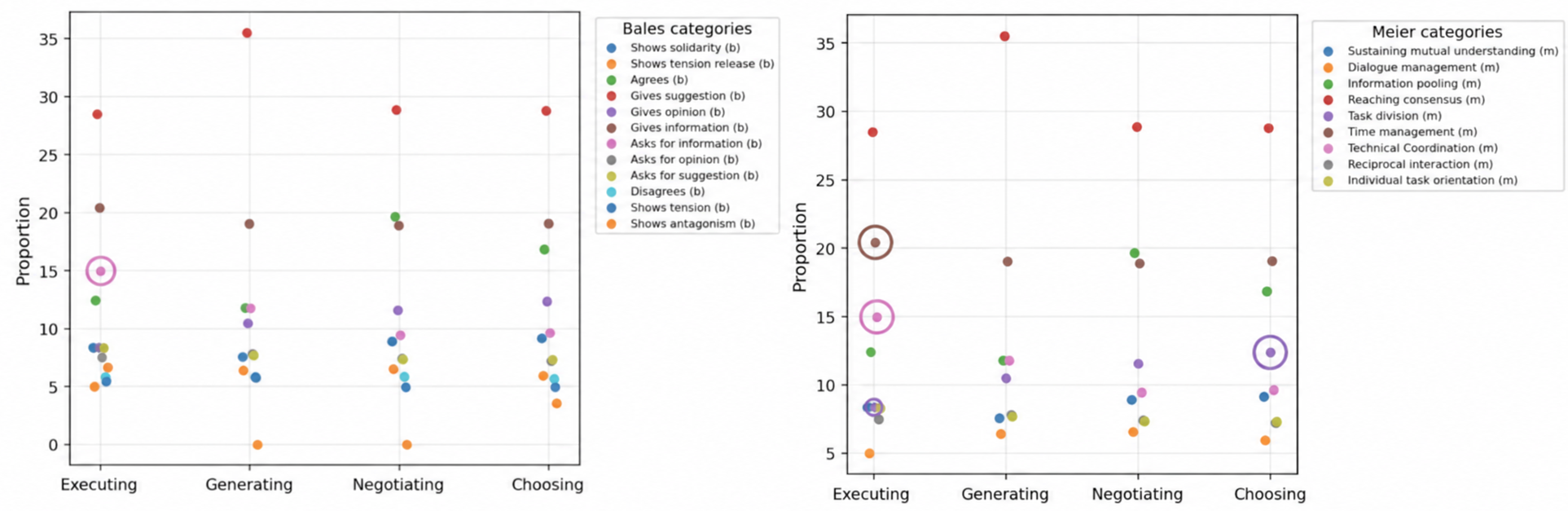}
    \caption{Effect of goal type on Bales labels and Meier collaboration dimensions for `scripted' simulations. Circled points indicate goal types that are significantly higher or lower for a given dimension after correction.}
    \label{fig:meier_goal_types}
\end{figure*}

Figure~\ref{fig:meier_goal_types} examines how collaboration behavior varies across goal types. Circled points indicate metrics that are significantly high or low for a given goal type after correction described in section \ref{sec:tukey}, with larger circles marking stronger deviations.

Compared with team size, goal type produces clearer and more interpretable behavioral differentiation. Execution and performance oriented tasks show elevated coordination related behaviors, especially time management and technical coordination. This suggests that simulators can respond to task structure by producing more procedural and coordination focused collaboration when the goal requires execution.

Problem solving and decision making tasks show stronger task division, indicating that agents adjust their collaboration style when the task requires evaluating options or dividing responsibilities. Ideation and planning tasks also produce high levels of suggestion and consensus related behavior, showing that the simulator adapts to generative task contexts by increasing proposal driven interaction.

However, goal type sensitivity is uneven. Negotiation and conflict resolution tasks do not strongly increase disagreement, tension, or antagonism in the overall interaction profile. This suggests that simulators are more sensitive to the task structure of collaboration than to its socially difficult dynamics. They can distinguish between execution, ideation, and decision making in terms of coordination and task organization, but they are less reliable at producing the friction expected in disagreement heavy settings.

Thus, goal type is the strongest contextual lever in our current results, but it mainly steers cooperative and task oriented behavior rather than conflict oriented behavior. 
In contrast, team composition did not exhibit
statistically significant effects on the measured
collaboration dimensions after multiple-comparison
correction. This suggests that, under our experimental
settings, varying expertise diversity has a weaker
influence on simulated collaboration behavior than
goal type or team size.

\subsection{Implications for Simulation Framework's Sensitivity and Utility}

Across results, we find that LLM based multi party simulation frameworks are sensitive to contextual factors, but this sensitivity is uneven. Framework design shapes the baseline interaction profile, team size effects depend strongly on the simulator, and goal type produces the clearest behavioral differentiation.

The strongest effects appear in task oriented collaboration. Simulated groups reliably exchange information, propose ideas, coordinate work, divide tasks, and build consensus. These patterns make the simulators useful for controlled studies of structured cooperation and for generating interaction data under varied task contexts.

The weakest effects appear in socially difficult aspects of collaboration. Disagreement, tension, negotiation pressure, and coordination breakdown remain muted across frameworks and goal types. This suggests that current LLM based multi party simulation frameworks may overproduce smooth collaboration and underrepresent the friction that often matters in real group settings.

Overall, our results support a qualified view of simulator utility. LLM based multi party simulation frameworks can be steered by context and framework design, making them useful experimental environments for studying some forms of collaboration. However, their behavior should not be treated as a faithful proxy for human group interaction without further validation, especially when the target phenomenon involves disagreement, conflict, or emergent coordination breakdown. 

\section{Conclusion}

We study the utility of LLM based multi party simulation frameworks for modeling collaborative group behavior. We frame simulation as a controllable experimental setting where contextual factors, including team composition, team size, goal type, and simulator framework, can be varied systematically to examine their effect on emergent collaboration. Using the pipeline introduced in Figure~\ref{fig:pipeline}, we construct teams from PersonaHub personas, instantiate them into collaborative scenarios, simulate interactions across multiple frameworks, and analyze the resulting conversations using behavioral metrics grounded in human collaboration research.

Our experiments show that LLM based multi party simulation frameworks are sensitive to contextual factors, but unevenly so. Across frameworks, simulated groups reliably produce task oriented collaboration, including information sharing, suggestions, coordination, and agreement. Goal type produces the clearest behavioral differentiation, while team size effects are more dependent on the simulation framework. We also find that framework design strongly shapes observed behavior: ``Scripted'' simulations often produce coherent and highly coordinated interactions, while more open ended frameworks such as Sotopia introduce greater interactional variability.

At the same time, current simulators underrepresent socially difficult aspects of collaboration, including disagreement, negotiation pressure, tension, and coordination breakdown. LLM group simulators are useful instruments for experiments around structured cooperation, but current systems may be unreliable for studying tasks/goals where group dynamics matter most (conflict, disagreement, and breakdown).


\section*{Limitations}

Our study is limited to simulated, rather than human, group interactions. Thus, we do not claim that LLM based simulators faithfully reproduce real teamwork. We mitigate this by grounding our analysis in established collaboration frameworks and by focusing on relative sensitivity to controlled contextual factors, rather than equivalence to human experiments.

Our experiments also use a fixed set of frameworks, factors, and goal types, with GPT-4.1 used for  simulation and GPT-5-chat used for annotation. This design helps isolate the effects of team size, composition, goal type, and framework choice, but limits generalization to other models, prompting strategies, and simulator designs. Extending the analysis to more models, human baselines, and richer organizational settings is left for future work.

Finally, our results show that current simulators underrepresent disagreement, tension, negotiation pressure, and coordination breakdown. We report this as a central limitation of current multi party simulators, especially for studying conflict heavy collaboration.

\section*{Ethical Considerations}

This work evaluates LLM based group simulators and should not be used to make claims about real teams, workplaces, or communities without human validation. Although our personas and scenarios are synthetic, simulated roles may still reflect stereotypes or biased assumptions. We therefore analyze aggregate collaboration behavior rather than individual or demographic traits, and frame our results as evidence about simulator sensitivity, not as a substitute for human social data.
\bibliography{anthology,custom,custom_2,gen}
\bibliographystyle{acl_natbib}

\appendix
\section{Related Work}
\label{sec:related}
\subsection{LLM-Based Simulation for Studying Behavior}

LLM-Based simulation has long been used to study how social patterns emerge from local interactions among agents \cite{schelling1971dynamic,epstein1996growing}. Recent LLM based simulation work extends this paradigm by using language models to generate human like responses, roles, and interaction patterns \cite{aher2023simulate,argyle2023outofone}. These approaches make simulation attractive for studying behavior at scale, but they also raise questions about validity: synthetic participants may not reliably substitute for humans, and their behavior can be sensitive to model and prompt design \cite{dillion2023ai,bisbee2024synthetic}.

This concern is central in recent user simulator work. Prior studies evaluate whether LLM based simulators maintain user goals, support long term collaboration, improve assistant training, match real user behavior distributions, or serve as reliable proxies for multi turn assistant evaluation \cite{mehri2025goal,mehri2026multisessioncollab,suh2026quantifying,dou2025simulatorarena}. Most of this work focuses on dyadic user assistant interaction. Our work extends this validation question to group simulation, where the target behavior is not only user realism but emergent collaboration among multiple agents.

\subsection{Multi Agent Social Simulation}

LLM-based simulators have been used to simulate rich social environments, role based interaction, cooperation, negotiation, and social intelligence. Generative Agents showed that memory based LLM-based simulators can produce believable social behavior in a sandbox environment \cite{park2023generative}. Other work studies collaborative reasoning, theory of mind, and open ended social interaction among agents \cite{li2023theory,zhou2024sotopia}. SOTOPIA and its extensions support interactive social evaluation and learning from simulated social interactions \cite{zhou2024sotopia,wang2024sotopia_pi,zhou2025sotopia_s4}.

Recent platforms make such simulations more configurable and scalable. GenSim provides a general social simulation platform \cite{tang2025gensim}, TinyTroupe supports persona rich multi agent simulation \cite{salem2025tinytroupe}, AgentSense evaluates social intelligence in interactive scenarios \cite{mou2025agentsense}, and ProMediate studies proactive agents in multi party negotiation \cite{liu2025promediate}. Recent work also shows that multi agent teams may fail to use expert members effectively, especially as team structure changes \cite{pappu2026experts}. These studies motivate our focus on whether contextual changes actually produce measurable changes in collaboration behavior.

\subsection{Contextual Factors in Collaboration}

Collaboration is known to depend on contextual factors such as team size, task type, role structure, expertise distribution, and communication setting \cite{mcgrath1984groups,hinde1987individuals}. Team research further shows that communication does not have a uniform relationship with performance across all contexts \cite{marlow2018team}. These findings motivate our central question: if LLM based multi party simulation frameworks are useful for studying collaboration, then factors such as team size, team composition, and goal type should systematically affect simulated collaborative behavior.

\subsection{Measuring Collaboration Behavior}

We draw on established collaboration coding frameworks to measure process level behavior rather than only task success. Bales' Interaction Process Analysis codes small group interaction into task oriented and socio emotional behaviors \cite{bales1950interaction}. Meier et al.'s collaboration quality scheme captures dimensions such as mutual understanding, dialogue management, information pooling, consensus formation, and coordination \cite{meier2007rating}. Prior work has used these frameworks to analyze multicultural team communication and computer supported collaborative problem solving \cite{nam2009process,kahrimanis2009assessing}. These frameworks are therefore well suited for evaluating whether simulated agents coordinate, negotiate, agree, disagree, and manage group interaction.

\subsection{Framework Design and Factorial Simulation}

Framework design and factorial contextual variation remain less studied in LLM based multi party simulation frameworks. Prior work shows that simulation design choices can strongly affect apparent social behavior \cite{zhou2024real}, while factorial designs can help identify which contextual factors drive simulated outcomes \cite{kalluri2026trust}. We address this gap by testing team composition, team size, and goal type across multiple simulation frameworks.

\begin{table}[t]
\centering
\small
\begin{tabular}{lcc}
\toprule
Ann. Task & Avg. Precision (\%) & Agreement (\%) \\
\midrule
Bales IPA & 84.1 & 72.7 \\
Meier Dim. & 100.0 & 100.0 \\
\bottomrule
\end{tabular}
\caption{Human validation of LLM judge annotations. Avg. Precision is computed as the average percentage of predicted labels marked correct by the two annotators. Agreement reports the percentage of overlapping completed items for which both annotators gave the same Yes/No judgment.}
\label{tab:llm_judge_validation}
\end{table}

\section{Manual Validation of LLM Judge Annotations}
\label{app:manual}
Following prior work on validating LLM based interaction classifiers \cite{mysore2025prototypical}, we conduct a manual precision based validation of our LLM judge predictions. Two co-authors, both familiar with Bales' IPA labels and Meier's collaboration dimensions, independently annotated a label balanced validation set using shared annotation guidelines. For Bales IPA, annotators judged whether the predicted utterance level IPA label correctly described the utterance. For Meier collaboration dimensions, annotators judged whether a predicted snippet level dimension was present in the snippet. Since multiple Meier dimensions can apply to the same snippet, each validation item presents one snippet and one predicted dimension, and asks annotators to judge whether that dimension is present.

Table~\ref{tab:llm_judge_validation} reports the validation results. The judge achieves 84.1\% average precision for Bales IPA and 100.0\% for Meier dimensions on completed annotations, with annotator agreement of 72.7\% and 100.0\%, respectively. Because Meier labels are multi-label and relatively dense, we interpret this result as a precision check rather than a full recall evaluation.


\noindent \textbf{Human validation question for Bales' IPA:}
\begin{tcolorbox}[before skip=0.3cm, after skip=0.3cm, middle=0.1cm, top=0.15cm, bottom=0.15cm, colframe=black]
\textit{Utterance:} $<utterance>$\\

\textit{Predicted IPA label:} $<ipa\_category>$\\

\textit{Question:}\\
\textit{Is this predicted IPA label correct?}\\
\texttt{[Yes / No]}\\

\textit{If No, what is the correct IPA label?}\\
$<12\_category\_dropdown>$
\end{tcolorbox}

\noindent \textbf{Human validation question for Meier collaboration dimensions:}
\begin{tcolorbox}[before skip=0.3cm, after skip=0.3cm, middle=0.1cm, top=0.15cm, bottom=0.15cm, colframe=black]
\textit{Conversation snippet:} $<snippet>$\\

\textit{Predicted Meier dimension:} $<meier\_dimension>$\\

\textit{Question:}\\
\textit{Is this predicted Meier collaboration dimension present in the snippet?}\\
\texttt{[Yes / No]}
\end{tcolorbox}

\section{Prompts for LLM-as-a-Judge}
\label{app:full_prmopts}
We use an LLM-as-a-Judge to extract structured collaboration annotations from simulated multi-agent conversations. The judge assigns utterance-level Bales IPA categories and segment-level Meier collaboration quality dimensions. We include the exact prompts below to make the annotation procedure transparent and reproducible.

\section{Other Implementational Details}
\label{app:responsible_research}
\subsection{Artifacts, Licenses, and Intended Use}
\label{app:artifacts}

Our study uses several existing scientific artifacts. We use PersonaHub \citep{ge2024scaling} as the source of synthetic persona descriptions, Sotopia \citep{zhou2024sotopia} and TinyTroupe \citep{salem2025tinytroupe} as LLM based multi party simulation frameworks, and OpenAI API models for scenario generation, conversation simulation, and LLM judge annotation. We cite the creators of these artifacts in the main paper and related work.

PersonaHub is released on Hugging Face under a CC BY-NC-SA 4.0 license. We use PersonaHub only for research purposes, consistent with the dataset's stated intended use. We do not redistribute the raw PersonaHub data. Instead, we use it to construct synthetic teams and generate aggregate simulation analyses. If we release any derived synthetic simulation outputs, we will do so only for noncommercial research use and subject to the attribution and share-alike conditions of the PersonaHub license, where applicable. PersonaHub's release notes also warn that the generated data may contain inaccuracies, unsafe content, or biases; accordingly, we filter non-collaborative persona clusters and avoid interpreting persona behavior as evidence about real people or demographic groups.

Sotopia and TinyTroupe are used as simulation software frameworks. Both are publicly released under MIT licenses. We use them to instantiate and run multi party simulations, and do not modify their licenses or redistribute their code except in accordance with their public release terms. Our scripted baseline is implemented by the authors and is used only as a controlled comparison point for framework design.

We access OpenAI models through the API for scenario generation, simulation, and annotation. OpenAI's terms assign ownership of outputs to the user to the extent permitted by law, while also making users responsible for their inputs, outputs, and the accuracy and appropriateness of downstream use. We therefore treat all generated conversations and judge annotations as synthetic research artifacts requiring validation, not as ground truth human behavior. We do not use the generated data to train deployed systems, profile real individuals, or make claims about real teams.

The artifacts created by this study consist of synthetic multi party conversations, configuration metadata, LLM judge annotations, aggregate collaboration metrics, and analysis code. These artifacts are intended for research on simulator sensitivity, collaboration measurement, and evaluation of LLM based multi party simulation frameworks. They should not be used for evaluating real workers, real teams, or real communities without additional human validation and appropriate ethical review.

\subsection{Data Documentation and Statistics}
\label{app:data_stats}

All conversations analyzed in this work are synthetically generated. Each simulation is defined by a configuration consisting of team composition, team size, goal type, and simulation framework. We evaluate $2$ team compositions, $4$ team sizes, $4$ goal types, and $3$ simulation frameworks, for $96$ controlled settings. For each setting, we generate up to $10$ scenarios ($10$ for scripted, $3$ for the other frameworks). Team size ranges from $3$, $5$, $7$, or $9$ agents.

The resulting data consist of multi turn conversations, utterance level Bales IPA labels, and snippet level Meier collaboration dimension labels. We do not use train, development, or test splits because the study is not a supervised model training task. Instead, all simulations are used for aggregate behavioral analysis and statistical testing. We report aggregate distributions, significant effects, and corrected pairwise comparisons in the main results.
\subsection{Computational Setup}
\label{app:compute}

We use closed API based LLMs for simulation and annotation. Specifically, GPT-4.1 is used for scenario generation and conversation simulation, and GPT-5-chat is used as the LLM judge for Bales IPA and Meier collaboration annotations. Because these are closed models, their exact parameter counts are not publicly available. We therefore report the model names, API based setup, and total scale of the experiment rather than parameter counts.

\subsection{Software and Statistical Testing}
\label{app:software_stats}

We use Python for preprocessing, simulation orchestration, aggregation, and statistical analysis. LDA-based clustering is used only to construct uniform and diverse teams from the persona pool. We compute Bales IPA distributions at the conversation level and Meier dimension presence percentages at the conversation level. These metrics are then used as dependent variables in factorial ANOVA and post hoc comparisons. We use a corrected significance threshold of $p \leq 0.0005$ to account for multiple comparisons across metrics, factors, and frameworks.

\subsection{Use of AI Assistants}
\label{app:ai_assistants}

AI assistants were used during the research and writing process for limited support editing and grammatical  assistance, and formatting. All substantive research decisions, experimental design choices, analyses, and claims were reviewed and finalized by the authors.

\clearpage
\noindent \textbf{Judge for Bales' IPA:}
\begin{tcolorbox}[
  breakable,
  width=\linewidth,
  before skip=0.3cm,
  after skip=0.3cm,
  middle=0.1cm,
  top=0.15cm,
  bottom=0.15cm,
  colframe=black
]
\textit{You are an expert in social psychology and communication analysis. Your task is to annotate a multi-person conversation using Bales' Interaction Process Analysis (IPA) framework.}

\textit{Background: Bales' IPA categorizes each utterance into one of 12 categories across 3 broader domains.}\\

\textit{A. Socio-Emotional Positive:}\\
\textit{- Shows solidarity: raises others' status, offers help, or shows appreciation $\rightarrow$ 1}\\
\textit{- Shows tension release: jokes, laughs, or expresses pleasure $\rightarrow$ 2}\\
\textit{- Agrees: shows understanding, acceptance, or affirmation $\rightarrow$ 3}\\

\textit{B. Task-Oriented Neutral:}\\
\textit{- Gives suggestion: proposes ideas, direction, or action $\rightarrow$ 4}\\
\textit{- Gives opinion: evaluates, analyzes, interprets, or expresses feelings $\rightarrow$ 5}\\
\textit{- Gives information: shares facts, data, or personal experience $\rightarrow$ 6}\\
\textit{- Asks for information: requests factual details or clarification $\rightarrow$ 7}\\
\textit{- Asks for opinion: seeks evaluation or perspective $\rightarrow$ 8}\\
\textit{- Asks for suggestion: seeks proposals for action or decision $\rightarrow$ 9}\\

\textit{C. Socio-Emotional Negative:}\\
\textit{- Disagrees: expresses dissent, doubt, or passive rejection $\rightarrow$ 10}\\
\textit{- Shows tension: displays frustration, anxiety, or concern $\rightarrow$ 11}\\
\textit{- Shows antagonism: criticizes, asserts dominance, or deflates others' status $\rightarrow$ 12}\\
\textit{- If none of the above categories apply $\rightarrow$ 13}\\

\textit{Label rules:}\\
\textit{- For each turn, assign the single most appropriate IPA category.}\\
\textit{- If a turn fits multiple categories, choose the category that best captures the speaker's main intent and social function.}\\
\textit{- Do not include explanations or any text outside the JSON output.}\\

\textit{Output format:}\\
\texttt{[}\\
\texttt{\ \ \{"turn": 1,
\\"ipa\_category": "4 - Gives suggestion"\},}\\
\texttt{\ \ \{"turn": 2,
\\"ipa\_category": "8 - Asks for opinion"\},}\\
\texttt{\ \ ...,}\\
\texttt{]}\\

\textit{Conversation:} \texttt{\{conversation\}}
\end{tcolorbox}

\clearpage
\noindent \textbf{Judge for Meier collaboration quality:}
\begin{tcolorbox}[breakable, width=\linewidth, before skip=0.3cm, after skip=0.3cm, middle=0.1cm, top=0.15cm, bottom=0.15cm, colframe=black]
\textit{You are an expert trained to assess collaboration quality using the Meier, Spada, and Rummel nine-dimension scheme.}

\textit{Read the following guidebook carefully and indicate which dimensions are present in the conversation segment. Use only evidence from the segment.}\\

\textit{Guidebook:}\\

\textit{- Sustaining mutual understanding: speakers make contributions understandable, explain terms, paraphrase, elicit feedback, give verbal feedback, demonstrate understanding, or ask for clarification $\rightarrow$ 1}\\
\textit{- Dialogue management: speakers maintain smooth communication flow, manage turn taking, avoid confusion about turns, signal handovers, or repair overlaps and misunderstandings $\rightarrow$ 2}\\
\textit{- Information pooling: partners collect solution-relevant information, relate new information to existing facts, explain relevance, elicit expertise, or clarify information needs $\rightarrow$ 3}\\
\textit{- Reaching consensus: partners critically discuss options, weigh arguments, challenge ideas, and finalize decisions based on facts $\rightarrow$ 4}\\
\textit{- Task division: partners split work into clear subtasks, plan steps systematically, match subtasks to expertise or resources, and divide work fairly $\rightarrow$ 5}\\
\textit{- Time management: partners monitor remaining time, schedule phases, evaluate discussion importance relative to time, and finish without rushing $\rightarrow$ 6}\\
\textit{- Technical coordination: partners coordinate technical tools, shared editors, copying and pasting, app switching, or parallel writing tools $\rightarrow$ 7}\\
\textit{- Reciprocal interaction: partners respect each other, encourage contributions, offer constructive critiques, share decision making, and contribute cooperatively $\rightarrow$ 8}\\
\textit{- Individual task orientation: participants actively pursue a good solution, focus on task-relevant information, avoid distractions, and show interest in high-quality work $\rightarrow$ 9}\\

\textit{Rating instructions:}\\
\textit{- For each of the 9 dimensions, decide whether it is present in the segment.}\\
\textit{- Return only the dimensions that are present.}\\
\textit{- Do not include explanations or any text outside the list.}\\

\textit{Output format:}\\

\texttt{[}\\
\texttt{\ \ "1 - Sustaining mutual understanding",}\\
\texttt{\ \ "6 - Time management"}\\
\texttt{]}\\

\textit{Conversation segment:}\\
\texttt{\{conversation\_snippet\}}
\end{tcolorbox}

\end{document}